\documentstyle[12pt,epsf]{article}
\newcommand{\spone}{1.1}

\newcommand{\singlespace}{\edef\baselinestretch{\spone}\Large\normalsize}

\title{Prospects for a  Quantum Dynamic  Random Access Memory (Q-DRAM)}
\author{S. Bandyopadhyay$\thanks{Corresponding author. E-mail: bandy@quantum1.unl.edu}$\\
\\Department of Electrical Engineering \\
University of Nebraska \\
Lincoln, Nebraska 68588-0511, USA}
\date{}

\begin{document}

\maketitle

\singlespace

\begin{abstract}

Compared to quantum logic gates, quantum memory has received far less 
attention. Here, we explore the prognosis for a solid-state, scalable
quantum dynamic random access memory (Q-DRAM), where the qubits are encoded
by the spin orientations of single quantons in exchange-decoupled quantum 
dots. We address, in particular, various possibilities for implementing
refresh cycles.

\end{abstract}


\pagebreak

\section{Introduction}

Quantum memory is an important constituent of quantum information 
science. It has many applications: (i) increasing the efficiency
of quantum key distribution (QKD) protocols (the receiver Bob 
stores the received qubits in a quantum memory and measures 
them {\it after} the sender Alice tells him the bases), (ii) improving
the EPR-based QKD schemes \cite{ekert}, (iii) teleporting a state using 
singlet pairs prepared in advance, (iv) new schemes for QKD 
that rely on the existence of short-term memory \cite{bennet, goldenberg},
(v) attacking oblivious transfer and quantum bit commitment schemes \cite{mor},
etc.

The requirements for quantum memory are thought to be 
very different from those of quantum 
gates.  In a quantum gate, the qubits are accessed 
and rotated numerous times, but the coherence time need 
not be very long; it simply has to be  
much longer than the switching time. In contrast, the qubits in a quantum 
memory are seldom accessed, but they must live much longer 
(ideally ``forever'') without decohering.
One must also be able to access them with high fidelity.

\section{Spintronic quantum memory}

The most popular scalable solid state quantum gates are based on manipulating the spins of single
electrons or holes in quantum dots \cite{loss, bandy}
or in  single dopant atoms \cite{kane, vrijen}.  For the 
sake of compatibility, we must implement quantum memory in the same systems$\footnote{The coherence times of electron or hole spins are much less 
than that of nuclear spins. However, nuclear 
spins are not easy to ``read'' as data; consequently, one must
couple the nuclear spin to an electron spin and then detect the 
electron spin \cite{kane} to read the original nuclear spin.
This transduction of a nuclear spin to an electron spin is a 
delicate process and difficult to implement with high enough fidelity.}$.

Because of the relatively short coherence time of electron or hole spins,
{\it non-volatile} quantum static memories (Q-SRAMs) are not
appropriate; rather, quantum
{\it dynamic memories} (Q-DRAMs) may be possible if the qubit can be {\it refreshed} 
periodically  through refresh cycles. Below, we explore possible 
routes to refreshing the quantum state of a quanton.

\subsection{Refreshing a qubit}

It is very possible that 
refreshing can be accomplished through the  {\it quantum
Zeno effect} which postulates that repeated observations 
 of a qubit will inhibit its decay \cite{sudarshan,
gurvitz}. Repeated observations automatically serve as  refresh cycles. However, this repeated observation has to be carried out
by a non-invasive detector. A ballistic point contact has been 
used in the past as a non-invasive charge detector for electrons 
in quantum dots \cite{field}, and its role in the
context of the quantum Zeno effect has been examined \cite{gurvitz}.
It may be possible to use a spin-polarized scanning tunneling
microscope tip as a non-invasive probe for spin, but that is yet to 
be realized in practice.

 A more straightforward approach 
would be to read the qubit periodically and then re-create {\it some}
(but not all) attributes of it. Since, we are not going to use the 
memory for computation (such as implementing Shor's or Kitaev's 
algorithms, or Grover's sorter), we may not need to draw upon
the full power of quantum parallelism. The expectations from 
``memory'' are different from those that we expect from ``logic gates''.
We may not need the full phase information in many cases.

Consider the qubit encoded by the coherent superposition of two 
spin states of a quanton:
\begin{eqnarray}
| \psi> = a_{\uparrow} |\uparrow> + a_{\downarrow} |\downarrow> \nonumber \\
|a_{\uparrow}|^2 + |a_{\downarrow}|^2 = 1
\end{eqnarray}
If we make several measurements of this qubit we will measure the 
upspin state with a probability $|a_{\uparrow}|^2$. Thus, by making 
several measurements over identical qubits, we can find the 
magnitudes of $a_{\uparrow}$ and $a_{\downarrow}$, but not the 
relative phase between them. {\it In many applications involving
measurements of stored qubits on given bases, it may be sufficient 
to know just the magnitudes of $a_{\uparrow}$ and $a_{\downarrow}$,
and the relative phase is unimportant}. For such niche applications,
we can develop a Q-DRAM with present technology as explained below.

One can periodically read the qubits (with a period much smaller than the
decoherence time) in several nominally 
identical hosts (e.g. single-electron quantum dots) to extract the magnitudes of $a_{\uparrow}$ and $a_{\downarrow}$. After each reading, we will re-inject quantons into these dots from
a spin polarized contact, followed
by immediate single qubit rotations in every dot to re-create the 
magnitudes of $a_{\uparrow}$ and $a_{\downarrow}$ (but not the 
relative phase, which will be arbitrary). In this fashion,
we can store the magnitudes of $a_{\uparrow}$ and $a_{\downarrow}$
for an indefinite time. This is a crude, but often effective, quantum dynamic random access memory (Q-DRAM).

\subsection{Why quantum dots?}

Single quantons confined in quantum dots are ideal storage media
from the vantage point of technology. Large, well-ordered arrays of quantum 
dots can be self-assembled with a density exceeding 10$^{11}$/cm$^2$.
This can lead to unprecedented density of qubits -- in excess of 
10$^{11}$/cm$^2$. Thus, even if we introduce a 100-fold redundancy (the same 
qubit is stored in 100 dots), we can still obtain a qubit storage
density in excess of 1 giga-qubit in 1 cm$^{2}$ (smaller than a postage stamp) whose storage 
capacity of 2$^{10^9}$ exceeds by far that of all the hard 
disks that could have been made with all the material in the 
universe over the life of the universe.

\begin{figure}
\epsfxsize=3.4in
\epsfysize=3.4in
\centerline{\epsffile{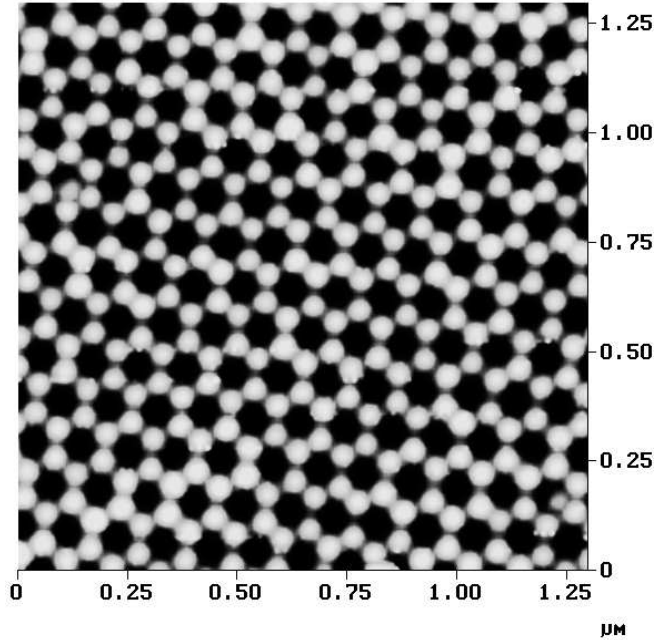}}
\caption[]{\small Raw atomic force micrograph of pore morphologies produced
by anodization of an aluminum foil  in oxalic acid. The average pore 
diameter is 52 nm with a 5\% standard deviation. This structure 
acts as a self-assembled template for self-assembling a quantum 
memory.\label{pores-top}}
\end{figure}

Fig. \ref{pores-top} shows a self assembled porous alumina film
produced in our laboratory by anodization of aluminum \cite{nanotechnology}. The pores 
can be filled sequentially with different materials to create 
multi-layered quantum dots surrounded by alumina. Alternately,
the pores can be used as etch masks to mesa-isolate quantum 
dots in a multilayered film grown by molecular beam epitaxy \cite{masuda}.
Using these techniques  , one can grow semiconductor quantum 
dots capped by ferromagnetic contacts that act as spin polarizers
for injecting spin, and spin analyzers for detecting spin.

\section{Quantum erasure}

Before concluding this paper, we point out that there is theoretically
an intriguing possibility of actually re-creating the entire 
qubit (including the phase) after it has been ``read''. i.e. after the spin analyzer has detected
the spin orientation. This involves  {\it quantum
erasure} \cite{hillary, scully, kwiat} as explained below.

Consider a quanton in a coherent superposition of two spin states,
described by a wavefunction
\begin{equation}
\psi = a_{\uparrow} |\uparrow> + a_{\downarrow} |\downarrow>
\end{equation} 

A fundamental result of quantum measurement theory is that if
the spin analyzer tries to detect the spin of the incoming quanton,
the wavefunction of the detector becomes entangled with that 
of the quanton. The entangled (non-factorizable) wavefunction
is
\begin{equation}
\Phi = a_{\uparrow} |\uparrow> |1> + a_{\downarrow} |\downarrow> |2>
\end{equation}
where the wavefunctions $|1>$ and $|2>$ span the Hilbert space 
of the detector. Thus, $|1>$ corresponds to the detector (spin-analyzer)
passing an up-spin quanton, and $|2>$ corresponds to the
detector reflecting a downspin quanton.

If we make a measurement of whether the detector passed the quanton
(corresponding to the determination that the quanton's spin was ``up''),
the probability amplitude of that is
\begin{equation}
\Psi = <1|\Phi> = a_{\uparrow} |\uparrow> <1|1> + a_{\downarrow} |\downarrow> <1|2>
\end{equation}
Since the detector makes an  ``unambiguous'' determination, meaning that 
it {\it always} passes an upspin quanton and {\it never} passes a
downspin quanton, the wavefunctions $|1>$ and $|2>$ are orthogonal,
meaning that upspin detection and downspin detection are mutually
exclusive (a quanton cannot be simultaneously both upspin and downspin,
and the detector will unambiguously determine what the spin is).
Hence, from Equation (4),
\begin{equation}
\Psi_{detected} = a_{\uparrow} |\uparrow>; ~~~~~ |\Psi|^2 = |a_{\uparrow}|^2
\end{equation}
and we get no information about $a_{\downarrow}$, or the phase. 
This is interpreted as wavefunction collapse. However, this collapse
is not irreversible since if we design an experiment whose 
result is the probability of a particular outcome of the spin measurement
{\it and} finding the detector in the symmetric state ($|1> + |2>$),
then the corresponding probability amplitude is 
\begin{eqnarray}
[<1| + <2|]|\Phi> & = & a_{\uparrow} |\uparrow> <1|1> + a_{\downarrow} |\downarrow> <1|2> + a_{\downarrow} |\downarrow> <2|2> + a_{\uparrow} |\uparrow> <2|1> \nonumber \\
& = & a_{\uparrow} |\uparrow> + a_{\downarrow} |\downarrow> \nonumber \\
& = & \psi ~,
\end{eqnarray}
which is the original wavefunction. Hence, we have restored the original 
wavefunction.

Note that if we can find the detector in the symmetric state, we would 
not have known whether the quanton that passed through it was ``up'' 
or ``down'', and hence we would not have collapsed the wavefunction.
Thus, by finding the detector in the symmetric state, we have 
erased the information about the spin and hence restored the 
original coherent superposition state.
The quantum erasure is possible because the entangled wavefunction 
$\Phi$ is still a pure state and not a mixed state.

\begin{figure}
\epsfxsize=4.3in
\epsfysize=3.4in
\centerline{\epsffile{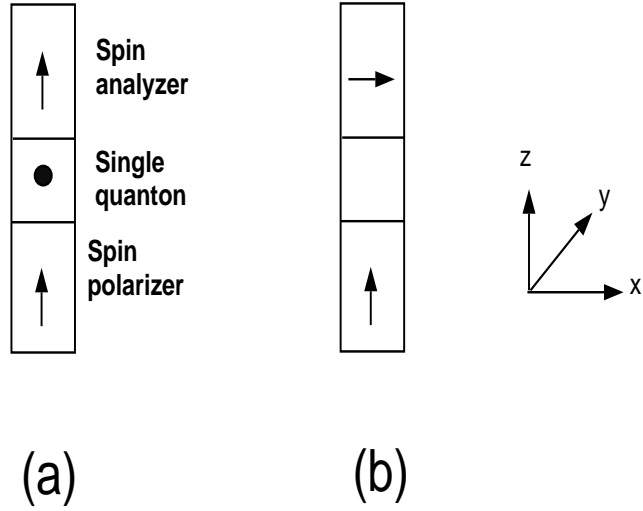}}
\caption[]{\small The (a) initial and (b) final state of the polarizer-analyzer combination after the passage of a quanton corresponding to the 
reading of a qubit. \label{erasure}}
\end{figure}

What do we need to implement quantum erasure? We need only one difficult technological feat. When the 
quanton passes through the spin analyzer, it should be able to 
rotate the magnetization of the analyzer and {\it change it}.
 If the
polarizer and analyzer were originally magnetized in the +z-direction, the 
passage of the quanton through the analyzer must turn on some 
interaction that results in the analyzer getting magnetized in the 
+x-direction. Fig. \ref{erasure} depicts this situation. We assume that 
$|1>$ corresponds to the state of the detector whereby the analyzer is 
magnetized (spin polarized) in the +z-direction and $|2>$ corresponds to the state of the 
detector whereby the analyzer is magnetized in the -z-direction. Thus,
\begin{equation}
|1> \rightarrow \left [ \begin{array}{c}
             1 \\
             0 \\
             \end{array}   \right]
\end{equation}
\begin{equation}
|2> \rightarrow \left [ \begin{array}{c}
             0 \\
             1 \\
             \end{array}   \right]
\end{equation}
Clearly $|1>$ and $|2>$ are orthogonal and the state $|1> + |2>$ 
corresponds to the state
\begin{equation}
|1> + |2> \rightarrow \left [ \begin{array}{c}
             1 \\
             1 \\
             \end{array}   \right] ~,
\end{equation}
which corresponds to spin-polarization in the +x-direction.

Thus, we must find the analyzer polarized in the +x-direction 
after the quanton passes through it.
In other words, the magnetization of the analyzer must be 
sensitive to the passage of a quanton and respond to it.
At present, this is not possible; but the recent discovery of 
control of magnetization via an electric current in InMnAs
\cite{ohno2} is beginning to hold out some hope in this direction.

\section{Conclusion}

In this brief report, we have pointed out the possibilities of quantum-dot
based quantum dynamic random access memory (Q-DRAM), and outlined three possible
schemes to implement qubit refreshing. The advantage of quantum dot 
based memory is the exceedingly large storage density that is possible.

\pagebreak


\begin{thebibliography}{10}

\bibitem{ekert}
A. Ekert. {\it Phys. Rev. Lett.}, {\bf 67}, 661 (1991).

\bibitem{bennet}
C. H. Bennet and S. J. Wiesner, {\it Phys. Rev. Lett.}, {\bf 69}, 2881
(1992)

\bibitem{goldenberg}
L. Goldenberg and L. Vaidman, {\it Phys. Rev. Lett.}, {\bf 75}, 1239 
(1995).

\bibitem{mor}
T. Mor, D.Sc. thesis, LANL e-print quant-ph/9900073.

\bibitem{loss}
G. Burkard, D. Loss and D. P. DiVincenzo,
{\it Phys. Rev. B}, {\bf 59}, 2070 (1999).

\bibitem{bandy}
S. Bandyopadhyay, {\it Phys. Rev. B}, {\bf 61}, 13813 (2000).


\bibitem{kane}
B. E. Kane
[{\it Nature}
(London), {\bf 393}, 133-137 (1998).

\bibitem{vrijen}
R. Vrijen, et. al., {\it Phys. Rev. A}, {\bf 62} 12306 (2000).




\bibitem{sudarshan}
B. Misra and E. C. G. Sudarshan, {\it J. Math. Phys.}, {\bf 18}, 756
(1977).

\bibitem{field}
M. Field, et. al, {\it Phys. Rev. Lett.}, {\bf 70}, 1311 (1993).


\bibitem{gurvitz}
S. A. Gurvitz, {\it Phys. Rev. B}, {\bf 56}, 15215 (1997).



\bibitem{nanotechnology}
S. Bandyopadhyay, et. al.,  {\it Nanotechnology}, Vol 7, 360 (1996);
S. Bandyopadhyay and A. E. Miller, in {\it Handbook of Advanced Electronic 
and Photonic Materials and Devices}, Vol. 6 (Nanostructured
Materials), Ed. H. S. Nalwa (Academic Press, New York, 2000), Chapter 1.

\bibitem{masuda}
H. Masuda and M. Satoh,  {\it Jpn. J. Appl. Phys.}, {\bf 35}, L126 (1996).

\bibitem{hillary}
M. Hillery and M. O. Scully, in {\it Quantum Optics, Experimental
Gravitation and Measurement Theory}, Eds. P. Meyestre and M. O. Scully
(Plenum, New York, 1983), p. 65.

\bibitem{scully}
M. O. Scully, B. J. Englert and H. Walther, {\it Nature} (London),
{\bf 351}, 111 (1991).

\bibitem{kwiat}
P. G. Kwiat, A. M. Steinberg and R. Y. Chiao, {\it Phys. Rev. A},
{\bf 49}, 61 (1994).

\bibitem{ohno2}
H. Ohno, et. al., {\it Nature} (London), {\bf 408}, 944 (2000).



\end{thebibliography}
\end{document}